# PLASMA LENS FOR US BASED SUPER NEUTRINO BEAM AT EITHER FNAL OR BNL*

A. Hershcovitch, W. Weng, M. Diwan, J. Gallardo, H. Kirk, B. Johnson,
Brookhaven National Laboratory, Upton, New York 11973, U.S.A.
S. Kahn, Muons Inc, Batavia, IL 60510, U.S.A.
E. Garate, A. Van Drie, N. Rostoker
University of California, Irvine, CA 92697, U.S.A.

*Abstract*

The plasma lens concept is examined as an alternative to focusing horns and solenoids for a neutrino beam facility. The concept is based on a combined high-current lens/target configuration. Current is fed at an electrode located downstream from the beginning of the target where pion capturing is needed. The current is carried by plasma outside the target. A second plasma lens section, with an additional current feed, follows the target. The plasma is immersed in a relatively small solenoidal magnetic field to facilitate its current profile shaping to optimize pion capture. Simulations of the not yet fully optimized configuration yielded a 25% higher neutrino flux at a detector situated at 3 km from the target than the horn system for the entire energy spectrum and a factor of 2.5 higher flux for neutrinos with energy larger than 3 GeV. A major advantage of plasma lenses is in background reduction. In anti-neutrino operation, neutrino background is reduced by a factor of close to 3 for the whole spectrum, and for energy larger than 3 GeV, neutrino background is reduced by a factor of 3.6. Plasma lenses have additional advantages: larger axial currents, high signal purity: minimal neutrino background in anti-neutrino runs. The lens medium consists of plasma, consequently, particle absorption and scattering is negligible. Withstanding high mechanical and thermal stresses in a plasma is not an issue.

## INTRODUCTION

In many areas of research involving charged particle beams, various methods of magnetic focusing have been employed to enhance the flux of charged particles from a divergent source such as a production target,[1] or to confine ions emerging from the cross-over region of an ion diode to betatron oscillation for propagation to a small target a few meters away.[2] The method of choice for focusing of high energy charged particles, produced in nanosecond to microsecond bursts, that need to be transported for a distance of a meter or more has been the use of azimuthal magnetic fields that pull the particles radially inward as a consequence of the Lorentz force. Large currents that are oriented along the desired flight path of the charged projectiles usually generate strong azimuthal magnetic fields. Therefore, devices with large axial current can be utilized as lenses.

*Work supported by Work supported under Contract No. DE-AC02-98CH1-886 with the US Department of Energy.
hershcovitch@bnl.gov

Lithium lenses [3,4] and horns [1,5] have been used in high energy physics research, while various spark, and Z channels were developed for fusion experiments.[2,6,7,8] Spark, Z channels, Z-pinches shall be referred to as plasma lenses, even though in high energy physics research this term was used for lithium lenses and lenses where lithium was replaced by high pressure gases.

Although some features vary from experiment to experiment, there are a number of common requirements including lenses for the Super Neutrino Beam:

1. Very large axial electrical currents (approaching a Mega-amp) must be generated and sustained.
2. The magnetic fields generated by these currents should capture the largest number of parent pions.
3. The lens medium should have lowest density possible to minimize pion absorption and scattering.
4. The lens must endure high mechanical and thermal stresses caused by pulsing high currents and EM fields.
5. Lens must survive prolonged exposure to radiation.
6. The lens should minimize neutrino background during anti-neutrino beam runs (signal purity).
7. A cost-effective, power-efficient lens is desirable.

For generating large neutrino beams, high-energy pions must be captured and maintained as a beam until they decay. Description and comparison of the various lenses, with focus on a plasma lens is presented in this paper.

## LENS OPTIONS

Interest in this type of charged particle focusing is varied and many applications require customized lens configuration. The lens choice in this paper, however, is done based on applicability to neutrino generation.

### *Horns*

A horn system is a hollow coaxial structure of conductors through which large currents (up to 300 kA) flow to generate the focusing magnetic fields.[1,5] Requirements on the inner horns are extremely demanding: they have to withstand very large thermal and mechanical stresses from pulsed operation, yet they must be fabricated from light elements to minimize particle losses. There is a limit to the current that can be carried. Additionally, horns do not capture pions with velocities that are at very small angles to target axis.

### *Lithium Lens*

A lithium lens consists of a lithium cylindrical conductor through which a large axial current is induced to generate an azimuthal magnetic field.[3,4] Unsuccessful attempts were made to replace lithium with compressed gases or aluminum. Lithium is contained under high pressure in a strong, chemically compatible, metal container.

### *Spark (or Z) channels*

Spark (or Z) channels are plasma transport channels, characterized by large currents (100s of kA), which have been developed to transport (and focus) intense beams of light ions over distances of up to 5 meters.[8] Channel radii from 1 cm to over 10 cm were reported (larger radii are easy to generate; radii below 1 cm are next to impossible).[8,9] Pulse lengths of 10s of nsec at a repetition rate of 500 Hz - 1 kHz have been generated, as well as 3 µsec long pulses at lower repetition rates.

These channels consist of two biased annular plates (or rings) placed in a vacuum chamber. The vacuum chamber is usually filled to a pressure of a few Torr to as high as 40 Torr with a gas. After an appropriate bias (10s of kV) is applied a spark or a laser pulse initiates a discharge that heats the gas. A large variety of these channels have been made, and an even larger variety is possible.[9] Another feature of these channels, which adds to their versatility, is the ease with which the direction of the discharge current can be changed.

### *Mega-Ampere Electron Beams*

Electron beam currents that are in the Mega-Ampere range have been generated by diodes. Although most of these diodes operate with pulses that are in the nsec range, some diodes have operated with pulse lengths of up to 2 microseconds. A hybrid system in which an electron beam is propagated through a plasma channel can be a very attractive option, since neither technique needs to be "pushed" to its technological limit to reach resultant axial currents exceeding 1 MA that are 1 meter long. Hollow-beam electron guns may be particularly suitable for such an application due to their larger perveance, enhanced stability, and their hollow structure.

### *Z-Pinches*

A Z pinch involves a sudden compression of low-density plasma by means of a large discharge current that lasts for a few microseconds. It bears some superficial similarity to a spark channel in that a discharge is formed between two end plates, but their plasma properties different, since the Z-pinch fill pressure is below a milli-Torr. In a series of experiments with magnetized Z pinches, 2 MA, 250 µsec were reached for a length of 0.8 meters.[10] Present day Z-pinch research involves discharge currents of 10 MA over a few centimeters.[11]

## NOVEL PION CAPTURE LENS

Presently, a horn system is being considered for pion capturing in the Super Neutrino beam. The first focusing lens is a 250 kA horn with an inner (outer) radius of 0.8 cm (8 cm) surrounding the 6-mm radius, 80 cm long carbon target.[12] A lithium lens is not an attractive alternative to a horn system since radius of a lithium lens is 1 cm or smaller,[13] the magnetic field at a distance of about 10 cm from the its axis (most important for focusing) is an order of magnitude lower than at the lens radius.

As an alternative, magnetized Z-pinches were first considered (can be flared), and since forty years ago a 1.5-meter long, 40-cm diameter Z-“pinch” lens, with a current of 500 kA for 15 µsec duration was successfully used in an AGS experiment.[14] This lens performed very well until its ceramic liner broke and was not replaced since the experiment was close to its conclusion.[15] Since then various special kevlar, fiberglass, and carbon epoxy liners and insulators (durable under extremely intense radiation) were developed for radiation generating machines.

Figure 1 is a display of lens/target configuration. Figure 1a is the 3-D embodiment, while 1b a schematic of the configuration. Part of the plasma straddles the target. Current is fed at an electrode located near the beginning of the target where pion capturing is needed. An additional current feed, at the end of the target facilitates higher (or different) current in the down stream part of plasma lens. The plasma lens is immersed in a solenoidal magnetic field to facilitate its current profile shaping.

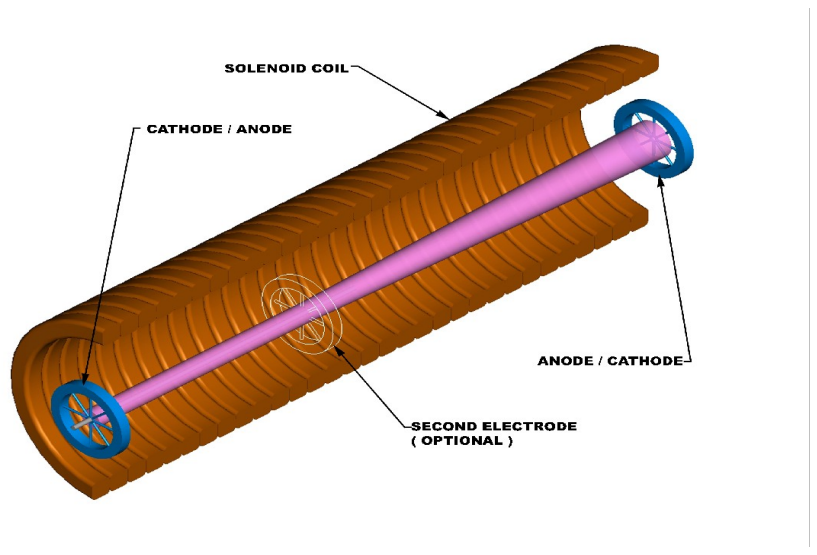

Figure 1a: Lens/target embodiment.

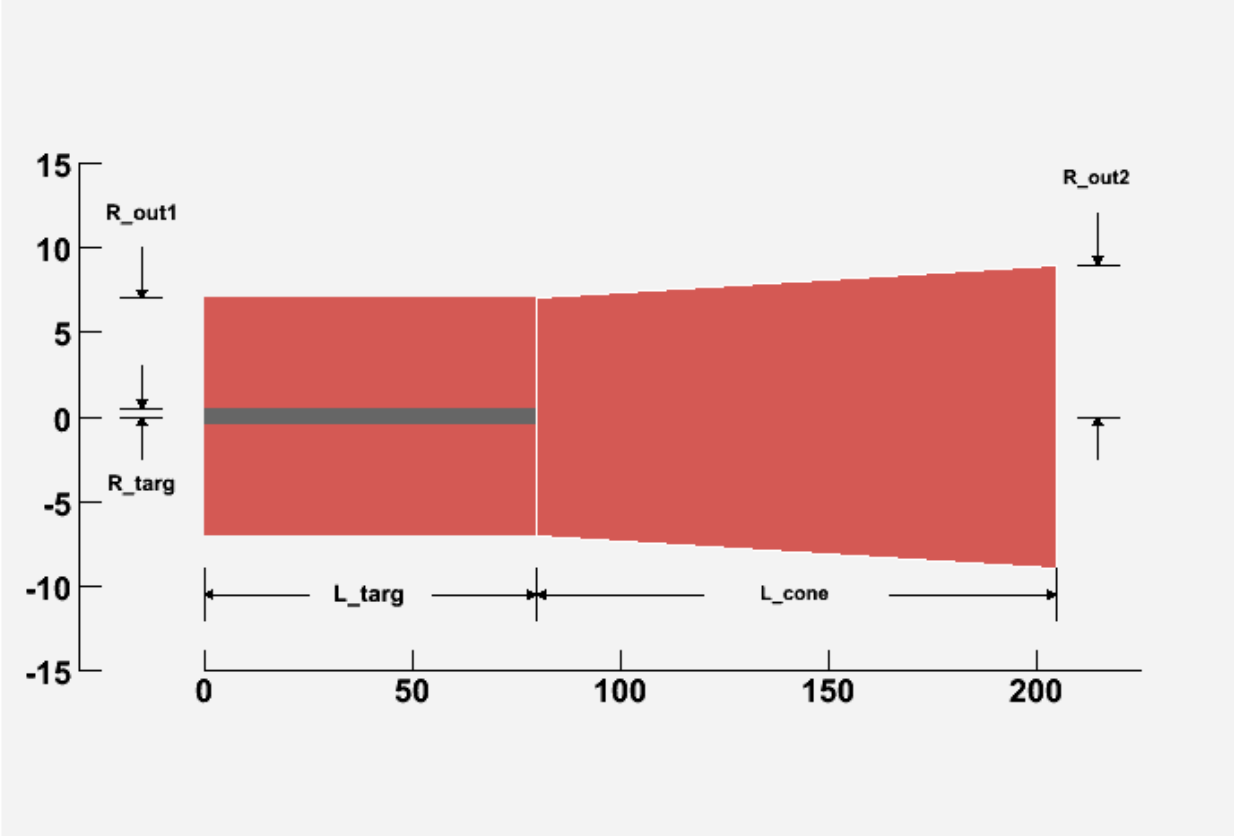

Figure 1b: Schematic of the plasma lens and target. Dimensions are given in cm.

Neutrino yield simulations for the above lens were performed for a 28 GeV proton beam on the carbon target. Range of simulated plasma outer radii was 3-12 cm for the section straddling the target, while outer radii of the flared section end was 5-15 cm. Basically, the outer radius of the “straight” plasma lens section R_out1 and the end of the “flared” section R_out2 were varied by the same amount, but R_out2 remained 3 cm larger than R_out1. In figure 1, the plasma is shown in pink. Carbon target (6 mm radius, 80 cm long[12]) is shown in gray. Plasma current was the same throughout the lens in all simulations. But, it is possible to flow different current in different sections of the lens, hence, the optional electrode in figure 1a.

Results of simulations are shown in figures 2 and 3 for neutrino flux and anti-neutrino to neutrino ratio at a detector 3 km from the target. Displayed in figures 2 and 3 are fluxes in neutrinos per $m^2$ per proton on target (Fig. 2) and neutrino/anti- neutrino ratios for various R_outer1 and plasma current values (in different colors) (Fig.3). Results are compared to those obtained for the horn[12] in a dashed line. The second lens is a down stream horn[12] unchanged from the original BNL design. Displayed results are for the whole neutrino spectrum (2a and 3a) as well as for neutrinos with energies larger than 3 GeV (2b and 3b).

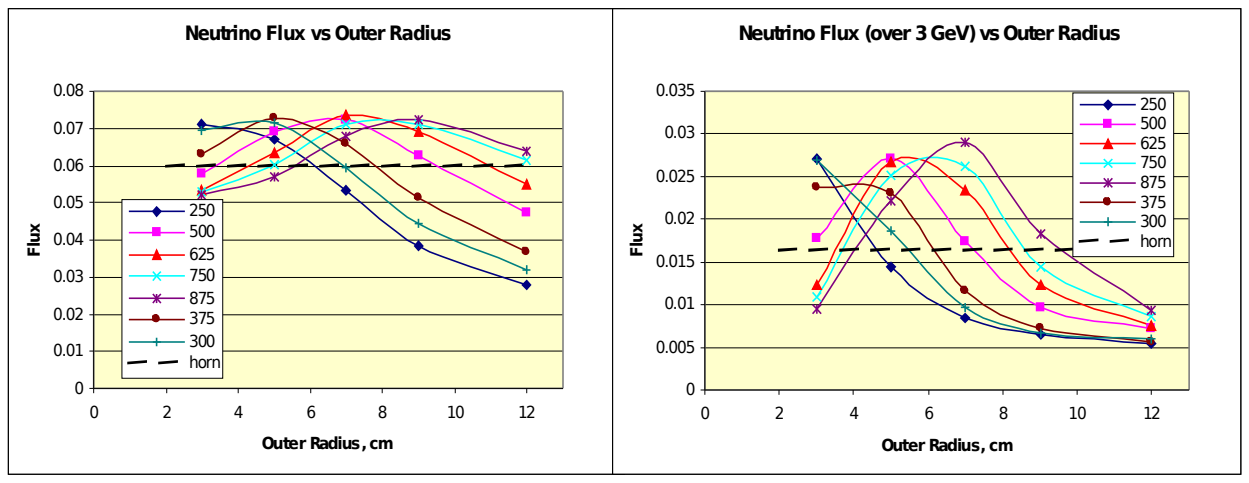

Figure 2: Neutrino flux vs. lens current and radius for whole energy spectrum (a), for greater than 3 GeV neutrinos (b).

Optimal overall neutrino flux is for plasma lens current of 375 and 625 kA for outer radii of 5 and 7 cm respectively, while high-energy (> 3GeV) neutrinos have optimal flux at currents of 250, 300, and 625 for outer radii of 3,5, and 7 cm respectively.

Comparing these neutrino fluxes to those obtained with horn as first pion focusing lens reveal a 25% overall neutrino gain in using a plasma lens. Results are more dramatic for high-energy neutrinos, where the gain is a factor of 2.5!

For neutrino background reduction during anti-neutrino runs, the results, (Fig. 3) are impressive: a factor of 3 for the whole publish or reproduce the published form of this manuscript, or others to do so, for the United States Government purposes.

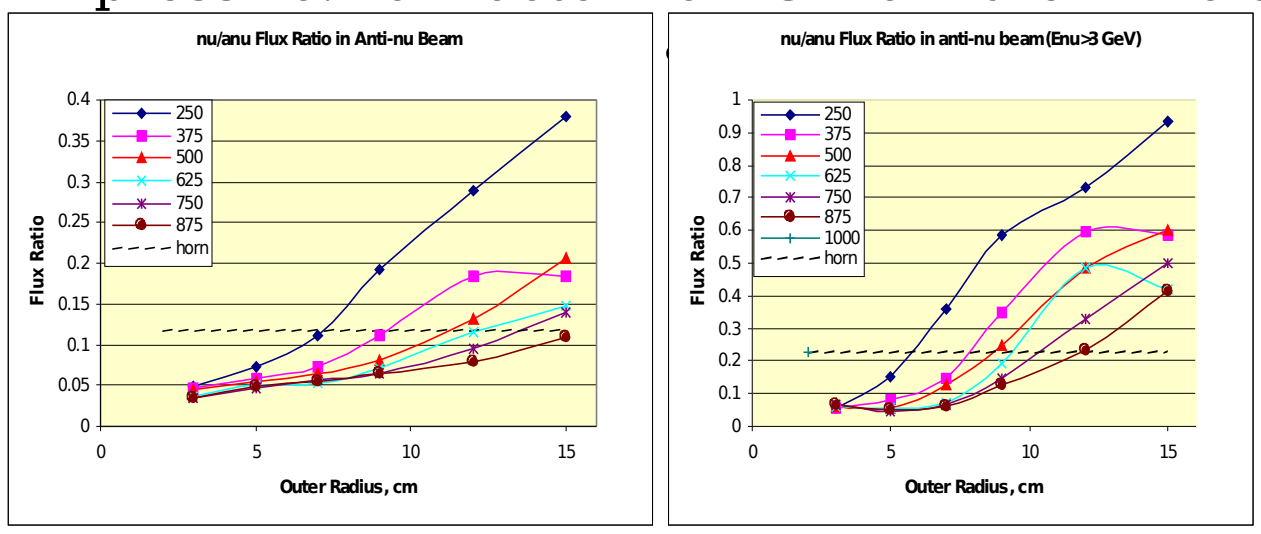

Figure 3: Neutrino to anti-neutrino ratio for whole neutrino spectrum (a); for neutrinos with energy larger than 3 GeV (b).

Optimal background reduction occurs for plasma lens radii of 3 – 7 cm, for almost any current.

## DISCUSSION

Large potential gain in neutrino flux coupled with very large reduction in background suggests that Z-pinches and spark channels deserve a further, more serious consideration. If further studies indicate that flaring a lens is beneficial, magnetized Z-pinch would be a better choice (adding a 1 kG magnetic field did affect the results). Conversely, a spark channel (no solenoid) might suffice if a straight lens is optimal. Additional optimization studies (of the flare and downstream lens) might yield further gains. Near term future studies include a similar study for a neutrino facility featuring 120 GeV proton beam at Fermilab.

## ACKNOWLEDGEMENT

Notice: This manuscript has been authored by Brookhaven Science Associates, LLC under Contract No. DE-AC02-98CH1-886 with the US Department of Energy. The U.S. Government retains, and the publisher, by accepting the article for publication, acknowledges, a world-wide license to publish or reproduce the published form of this manuscript, or others to do so, for the United States Government purposes.